%% file: main.tex
\documentclass[letterpaper,twocolumn,10pt]{article}
\usepackage{usenix}

\usepackage{tikz}
\usepackage{amsmath}
\usepackage{pifont}
\usepackage{mathrsfs}
\usepackage[normalem]{ulem}
\usepackage{graphicx}
\usepackage{multirow}
\usepackage{float}
\usepackage{framed}

\usepackage{listings}

\lstdefinestyle{bulkheadappendix}{
  basicstyle=\ttfamily\scriptsize,
  breaklines=true,
  breakatwhitespace=false,
  columns=fullflexible,
  keepspaces=true,
  showstringspaces=false,
  frame=single,
  framerule=0.3pt,
  framesep=3.5pt,
  backgroundcolor=\color{black!2},
  rulecolor=\color{black!35},
  xleftmargin=0.5em,
  xrightmargin=0.5em,
  aboveskip=5pt,
  belowskip=5pt,
  captionpos=b
}
\usepackage{diagbox}
\usepackage{threeparttable}
\usepackage{booktabs}

\usepackage{tabularx}
\usepackage{array}

\newcolumntype{L}[1]{>{\raggedright\arraybackslash}p{#1}}
\newcolumntype{Y}{>{\raggedright\arraybackslash}X}

\newcommand{\ignore}[1]{}
\definecolor{Blue}{RGB}{22,97,171}

\definecolor{RevisionBlue}{RGB}{0,0,0}

\begin{document}

\date{}
\title{Bulkhead: Automated Semantic Detection and Remediation of Container Escape Vulnerabilities}

\author{
{\rm Qiyuan Fan\textsuperscript{1},
Zhi Li\textsuperscript{1},
Junjie Li\textsuperscript{2}}\\
{\rm XiaoFeng Wang\textsuperscript{2},
Bin Yuan\textsuperscript{1},
and Deqing Zou\textsuperscript{1}}\\[0.5em]
\textsuperscript{1}Huazhong University of Science and Technology,
Wuhan, China\\
\textsuperscript{2}Nanyang Technological University, Singapore
}

\maketitle

\begin{abstract}
Container filesystem isolation depends on safely resolving container-controlled paths during privileged host-container interactions. As third-party plugins and agent sandboxes introduce more interactions, unsafe path resolution can redirect host operations across isolation boundaries, expanding the attack surface for \textit{path traversal} vulnerabilities. These logic flaws are difficult to mitigate: kernel defenses require intrusive redesign, static rules miss cross-boundary semantics, and manual auditing cannot scale. Large Language Model (LLM) based methods can detect syntax flaws through stable code patterns and identify logic flaws in domains such as smart contracts by comparing equivalent implementations. Container components, however, provide neither form of regularity: their heterogeneous interfaces operate under trust boundaries, filesystem views and interaction orders, making the security invariant difficult to find.


We present \textit{Bulkhead}, an automated framework for detection and remediation. Bulkhead summarizes historical CVEs into structured semantic patterns, correlates documentation with entry and effect side code evidence, and recovers source-to-sink execution contexts. Grounded patterns guide LLM reasoning, while isolated Proof-of-Concept (PoC) execution validates findings. For remediation, Bulkhead generates repository-specific patches, checks that they build, and iteratively verifies path safety and functionality using pattern derived model checking templates and counterexamples before runtime validation. Across container repositories, Bulkhead discovered 20 zero-day vulnerabilities, accounting for 43\% of the total number discovered over the past decade. Among them, 3 have been assigned CVEs and the verified patch suggestions have been submitted to the upstream.


\end{abstract}

\input{1-introduction}

\input{2-background}

\input{3-detection}
\input{4-patch}
\input{5-evaluation}

\input{6-relatedwork}

\input{7-conclusion}




\appendix

\input{8-appendix}

\bibliographystyle{plainurl}
\bibliography{ref}



\end{document}

%% file: 1-introduction.tex
\section{Introduction}
\label{sec:intro}

Container technologies leverage kernel mechanisms to provide lightweight virtualization and isolated execution environments for diverse workloads~\cite{jarkas2025container, Zhou2026Optimization, chen2026break}. As containers increasingly support AI workloads, they require direct access to specialized hardware, driving the adoption of third-party plugins for capabilities such as GPU acceleration. Containers are also becoming a foundation for autonomous agents; for example, systems such as OpenClaw~\cite{openclaw} use container-based sandboxes to isolate agent workspaces.

Despite these advances, logic vulnerabilities particularly container path traversal remain a persistent and growing threat. Such flaws have been disclosed almost every year over the past decade, resulting in 47 high-severity CVEs. They typically arise when privileged host processes insufficiently validate file paths influenced by a container. By manipulating symbolic links (symlinks), an attacker can redirect host-side operations to sensitive resources outside the container's isolation boundary.
Third-party plugins amplify this risk because their developers may overlook security checks required to preserve container isolation. For example, the NVIDIA Container Toolkit has recently disclosed multiple path-traversal vulnerabilities, including CVE-2024-0132~\cite{cve-2024-0132} and CVE-2025-23267~\cite{CVE-2025-23267}. Similar flaws have also compromised prominent container-based agent sandboxes, as illustrated by CVE-2026-31990~\cite{CVE-2026-31990} and CVE-2026-46703~\cite{CVE-2026-46703}.


\vspace{3pt}\noindent\textbf{Challenges in Mitigating the Security Risks.}
A complete protection against these vulnerabilities would require redesigning container isolation at the kernel level. However, such approaches typically demand intrusive modifications to core kernel components~\cite{li2023lost}, creating substantial barriers to adoption and upstream integration into the Linux mainline.

As a result, the container community largely relies on case-by-case discovery and patching of path traversal vulnerabilities. Detecting these logic flaws is particularly challenging because they lack clear syntactic signatures and instead require reasoning about the semantics of container operations and host–container interactions. This semantic complexity limits the effectiveness of traditional static analysis~\cite{chen2019research} and leaves manual auditing as the primary defense. Yet as the container ecosystem expands to include third-party plugins and containerized agent sandboxes, the resulting code volume and diversity make manual auditing increasingly impractical. Under this pressure, some container systems, such as Pouch~\cite{pouch}, simply disable risky filesystem operations altogether, sacrificing functionality for security.


\vspace{3pt}\noindent\textbf{LLM-based Logic Flaw Detection.}
A promising direction to address the above challenges is to leverage semantics-aware LLMs, which are increasingly used for vulnerability analysis. Syntactic vulnerabilities, such as buffer overflows, often expose local code evidence, such as dangerous APIs, source–sink relations, vulnerable snippets, or vulnerable–patched pairs, which allows a model to recognize recurring code and data-flow patterns \cite{Ahmed2025LLMxCPG}. Logic flaws are harder because they lack stable code-level signatures. In constrained domains such as smart contracts, however, explicit entry points and multiple implementations of the same protocol provide a useful basis for comparison, enabling models to infer missing checks from deviations across peer implementations \cite{gao2026logicscan}.

Container path traversal vulnerabilities present a harder setting: they exhibit neither stable code patterns nor a reliable basis for cross-implementation comparison (Section~\ref{subsec:challenges}). 
These vulnerabilities initially appeared in core container operations like volume mounting and file copying. As the ecosystem expands, the same isolation failure propagates to diverse extensions, such as GPU toolkits and containerized AI sandboxes. Serving distinct business logic, these functionalities manipulate file paths through different API abstractions and control flows, leaving no reusable code structure for detection. 
Furthermore, determining whether these path manipulations are vulnerable strictly depends on the complete execution context.
For example, Docker shares a unified copy implementation for both copies within the container and copies between the host and the container. Yet, only the latter requires a privileged host process to resolve an untrusted path within the host filesystem view. As the static code remains constant across these different contexts, analyzing the code structure alone cannot reveal potential flaws. Consequently, reliable detection of these logic flaws requires shifting the paradigm from static syntax matching to semantic reasoning over the complete execution context.

\vspace{3pt}\noindent\textbf{Our Approach.}
To overcome these limitations, we propose \textit{Bulkhead}, an automated LLM-based pipeline for end-to-end vulnerability detection and remediation.
For detection, Bulkhead summarizes historical CVEs into semantic patterns that capture isolation targets, their formalized execution contexts, and attacker capabilities.
These patterns represent the shared isolation failure independently of project-specific APIs. Guided by the patterns, Bulkhead reconstructs their complete security context, including container-controlled inputs, path transformations, security checks and resolving principals. The LLM then maps the pattern in this context to find whether there is an illegal operation. A candidate is reported only when there are boundary violations in the test for a PoC.
For remediation, the LLM generates a repository-specific patch from the validated finding. Since the patch may retain a check-use window or reject legitimate paths, Bulkhead performs formal verification using a model checking template abstracted from historical vulnerability patterns. Any counterexample returned to the LLM guides the next patch attempt. This generation and verification loop continues until the patch satisfies the specified path-safety and compatibility properties.



Bulkhead analyzed 82 container-related repositories and identified 20 zero-day vulnerabilities across 18 projects (22.0\%), enabling attacks such as sensitive host-file compromise and container escape (Section~\ref{subsec:discovery}). These findings amount to 43\% (20/47) of all publicly reported container path-traversal vulnerabilities from the past decade. We disclosed the vulnerabilities to the affected parties, all of whom acknowledged the significance of the discovered logic flaws; three CVEs have been assigned so far. We also submitted Bulkhead-generated and verified patches to upstream maintainers for review.

\vspace{3pt}\noindent\textbf{Contributions.}
Our contributions are outlined as follows:

\vspace{2pt}\noindent$\bullet$\ \textit{Persistent Threats and Attack Surface Expansion}
Our findings demonstrate that container path traversal remains a persistent threat, with its attack surface actively expanding from traditional container utilities into third party toolkits and containerized AI sandboxes.


\vspace{2pt}\noindent$\bullet$\ \textit{LLM-driven Logic Flaw Detection and Remediation}. We establish the core design principles for LLM reasoning over logic vulnerabilities. We demonstrate that extracting semantic patterns and supplying the complete execution context are critical to detect flaws across different target codebases. Bulkhead implements these principles into an automated pipeline for detection and patching.


\vspace{2pt}\noindent$\bullet$\textit{~Real-World Security Impact}. 
Evaluating 82 mainstream repositories, Bulkhead uncovered 20 zero day vulnerabilities that enable critical attacks including arbitrary host file access and complete container escape. This large scale discovery represents 43\% of the publicly reported container path traversal flaws over the past decade.


%% file: 2-background.tex
\section{Background}
\label{sec:background}

\subsection{Container Filesystem Isolation}
Container systems rely on namespaces \cite{biederman2006multiple} and cgroups \cite{cgroup} to isolate resources. To isolate the filesystem of a container, a mount namespace decouples its directory hierarchy from the host environment, and \texttt{chroot} confines its root path.
These mechanisms create an independent filesystem view that prevents internal container processes from viewing external host files. However, these restrictions do not achieve true Virtual File System (VFS) isolation in the kernel, where the container and the host continue to share a single VFS instance.
When a host process (e.g., container tools) accesses a file path inside a container, the kernel performs a VFS lookup operation without the ability to distinguish whether the resolved file object belongs to the container or the host.
This lack of boundary distinction prevents the kernel from enforcing effective access control, thereby allowing a malicious container to mislead the host process using symlinks and achieve a container escape~\cite{li2023lost}.
This limitation shifts responsibility for filesystem isolation to user-level container tools such as Docker and Podman, which establish the container's root directory. Therefore, they must identify and enforce access boundaries during host-container interactions, as inadequate enforcement introduces filesystem isolation vulnerabilities.



\subsection{Filesystem Isolation Vulnerabilities}

Filesystem isolation vulnerabilities arise during two container lifecycle stages. In the pre-start stage, the container engine constructs the root filesystem from an image, configures volumes or build directories, and executes hooks that inject devices or libraries. These operations precede the initial container process but run with host privileges. Their paths may derive from untrusted image content or user-provided configuration. MisResolution occurs when a privileged component lacks path validation, and resolves a container-controlled path in the host filesystem view. A malicious symlink or mount point can then redirect the privileged operation to a host object outside the container root.

\begin{figure}[t]
	\centering
	\includegraphics[width=.48\textwidth]{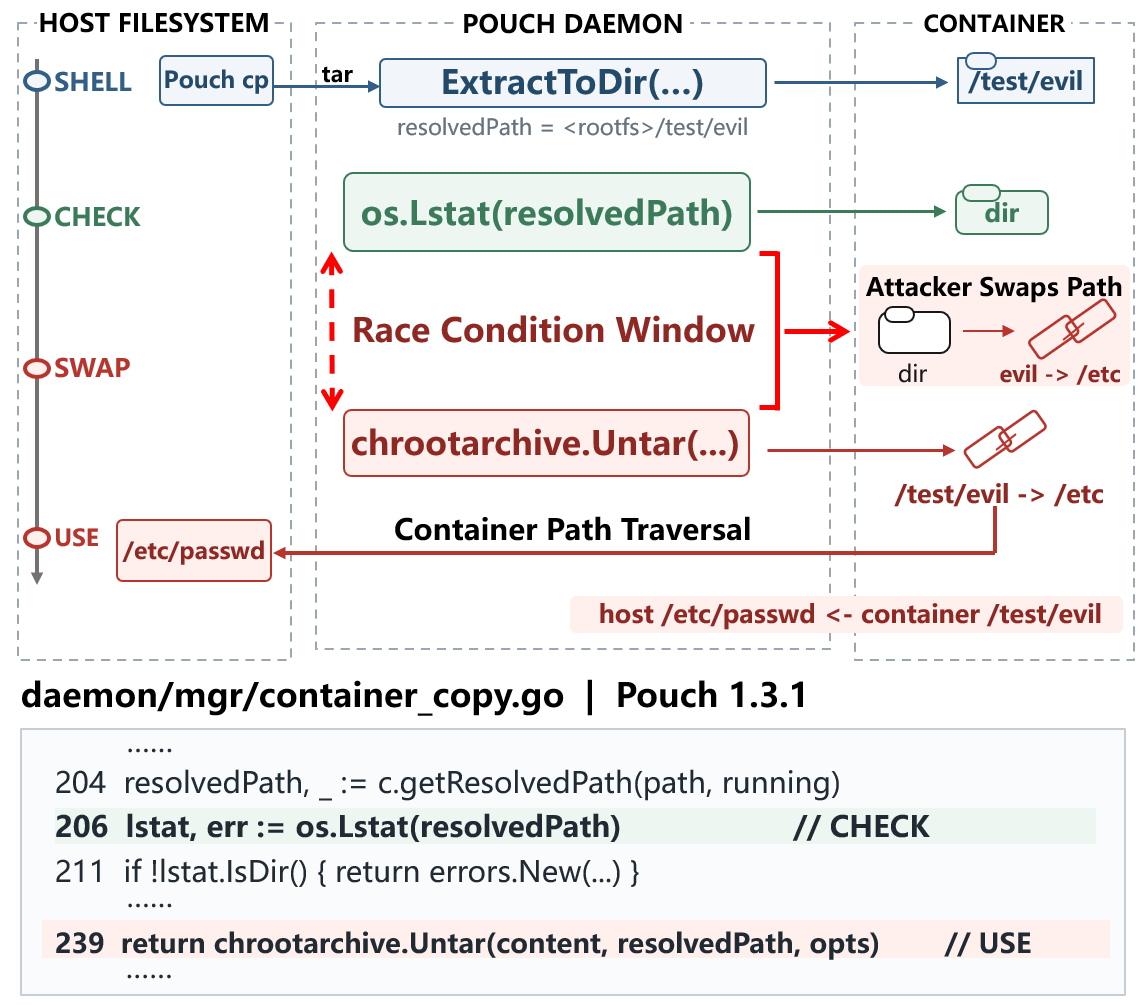}
	\caption{A typical container path traversal vulnerability.} 
	\label{fig:pouch-cp}
\end{figure}

The runtime stage begins when the initial container process starts. Privileged host daemons can perform file copying, archive extraction, and volume management on behalf of the container. A running container can modify directory entries while these operations execute. TOCTOU arises when validation and use resolve the same pathname at different times. The path identifies a legitimate object when checked, but the container can replace the checked path with a malicious symlink before use. The second resolution will then follow the symlink to an unvalidated host object. Pouch's copy implementation in Figure~\ref{fig:pouch-cp} instantiates this race (i.e., CVE-2024-4XX28, anonymized for ethical reasons). The operation \texttt{os.Lstat()} first confirms that \texttt{resolvedPath} refers to a legitimate directory. The attacker then replaces \texttt{/test/evil} with a malicious symlink to \texttt{/etc} before \texttt{chrootarchive.Untar()} resolves the path again. The extraction consequently follows the symlink into the host filesystem, leading to container escape.

During the past decade, container path traversal vulnerabilities have repeatedly occurred 47 times. Early cases were concentrated in container-management functions such as copy, mount, and archive extraction \cite{li2022rund,haq2024sok}. The attack surface now includes build systems, the container toolkit, and container-based AI sandboxes such as NVIDIA container toolkits, OpenClaw. They introduce shared-volume mounting, library injection, and workspace synchronization. These operations implement the same unsafe host-container interaction through different APIs, languages, and components. Code patterns tied to earlier functions therefore provide poor coverage of the expanded attack surface.


\subsection{Vulnerability Mitigation Limitations}
The recurrence of container path traversal vulnerabilities over the past decade shows that current approaches still have limitations. On the one hand, kernel-level enforcement can express strong isolation rules at VFS path lookup, but it requires intrusive kernel changes and may interfere with legitimate host-container operations. Therefore, the prior kernel-level design has not been accepted into the Linux mainline \cite{li2023lost}, limiting its deployment in existing systems. It also does not provide component-level remediation for path handling flaws in user space extensions and ecosystem tools. These constraints leave the expanding user-space attack surface without a practical repair path.

On the other hand, repository level case-by-case detection and remediation in community remains vulnerable to code evolution. CVE-2019-19921 exposed a race condition in volume mounting and was subsequently fixed \cite{cve-2019-19921}. While addressing the symlink traversal reported as CVE-2021-30465, developers changed the path-resolution workflow and inadvertently reintroduced the same race \cite{cve-2021-30465}. This regression remained undetected until it was disclosed as CVE-2023-27561 \cite{cve-2023-27561}. This sequence shows that although an individual detection and remediation may close the immediate attack, it may invalidate an invariant established by an earlier fix. Without an explicit security invariant preserved across components, such regressions remain difficult to prevent.

\begin{figure*}[t]
	\centering
    \includegraphics[width=0.98\textwidth]{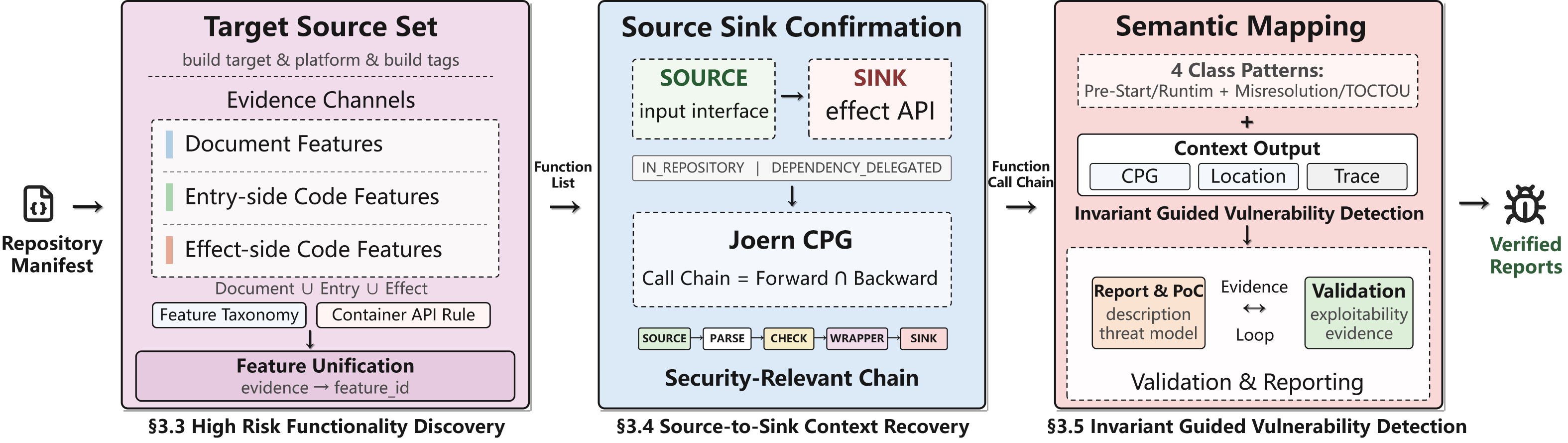}
	\caption{Bulkhead Detection Pipeline. 
    } 
	\label{fig:detection}
\end{figure*}

\subsection{LLM-based Vulnerability Detection}
The limitations above motivate the use of LLMs for semantic vulnerability analysis. Compared with fixed rules, LLMs can interpret cross-function context and the intent of security intention \cite{lin2025large}. A common workflow provides the model with vulnerability knowledge and selected code context, then asks whether the implementation violates a security condition. Its effectiveness therefore depends on how this knowledge and context are constructed \cite{wu2024lemur}.

Among existing designs, representative forms of vulnerability knowledge include syntax patterns and implementation contrasts. Syntax flaw detection commonly uses dangerous APIs, source-to-sink relations, vulnerable snippets, or vulnerable-patch pairs. These artifacts encode local code-level patterns and are effective when a flaw retains a stable implementation signature. Logic-flaw detection can instead infer required checks from comparable implementations. LogicScan groups smart contracts implementing the same protocol, starts from explicit \texttt{public} and \texttt{external} functions, and identifies deviations from behavior shared by peers \cite{gao2026logicscan}. Both forms rely on reusable regularity, either a stable code pattern or a reliable set of comparable implementations.




%% file: 3-detection.tex
\section{LLM-Driven Logic Flaw Detection}
\label{sec:detection}

\subsection{Challenges of Detecting Logic Flaws}
\label{subsec:challenges}

\noindent\textbf{Challenge I: Generalizing Logic Flaw Patterns}. 
A specific class of logic vulnerabilities shares common underlying principles but manifests in divergent concrete forms across distinct software components. These components possess diverging functional roles and execution contexts.
For example, earlier container path traversal vulnerabilities concentrated in copy, mount, and archive extraction functions \cite{li2022rund, haq2024sok}. The attack surface now expands to resource injection by the NVIDIA Container Toolkit, shared volumes, and workspace synchronization in AI sandboxes~\cite{CVE-2025-23359, CVE-2026-7875}. These target operations span different software stacks, lifecycle stages, APIs, and privilege arrangements. 
However, existing LLM-based detection pipelines target syntax vulnerabilities by reasoning based on known structural patterns found in abstract syntax trees, control flow graphs, API traces, or retrieved vulnerable code examples~\cite{Ahmed2025LLMxCPG}. This approach proves ineffective for logic flaws where a shared underlying principle results in heterogeneous code structures. The challenge is to construct a high-level representation that captures the core logic defect independently of specific component roles and language syntax.


\vspace{3pt}\noindent\textbf{Challenge II: Extracting Execution Context}. 
Logic flaws emerge when shared code spanning distinct execution environments fails to enforce the varying security requirements imposed by a specific context. For example, Docker file transfer operations reuse the similar underlying copy routines for operations confined within a container and for operations crossing the host container boundary. A copy command executed inside a container resolves paths within an isolated filesystem view \cite{dockercp}. In contrast, transferring files between the container and the host requires a privileged host component to resolve a container controlled path, introducing the risk of path traversal. Although both scenarios execute the same core copy logic, they rely on different resolving principals and security assumptions. Without execution context, analyzing isolated code fragments fails to determine whether critical security semantics are missing. Detecting logic flaws requires recovering a broader context that preserves external data sources, path transformations, privilege assignments, and final filesystem effects. Omitting any of these context relations obscures the security judgment.


\vspace{3pt}\noindent\textbf{Challenge III. Aligning Abstract Patterns with Concrete Execution Contexts.}
Unlike syntax vulnerabilities, logic vulnerabilities lack fixed code-level structural characteristics. Their detection depends on matching high-level abstract vulnerability patterns against concrete code implementations. However, feeding high-level vulnerability patterns and target code directly to LLMs for vulnerability detection is inherently unstable. The dimensional mismatch between abstract patterns and concrete code prevents LLMs from reliably mapping semantic roles to specific code instances. Resolving this gap requires explicit mechanisms to translate abstract vulnerability patterns into concrete execution contexts. The analysis must identify corresponding code elements and verify that their relations satisfy object identity and temporal constraints.


\subsection{Detection Pipeline}
\label{sec:detection_overview}
To address those challenges, Bulkhead provides an automated semantic pipeline for detecting container path traversal vulnerabilities across heterogeneous implementations. It first represents prior vulnerability knowledge as patterns that capture path interaction, object control, filesystem views, and attacker capabilities, which can be generalized beyond specific code structure, rather than project-specific code snippets (Section~\ref{subsec:detection}). During repository analysis, these patterns serve as security specifications and are grounded in concrete program evidence. Fig.~\ref{fig:detection} shows the workflow from repository manifests to validated vulnerability reports.

To extract the context, Bulkhead scopes the selected build to derive the Target Source Set and correlates repository documentation with entry-side and effect-side code evidence. The LLM maps project terminology to the Feature Taxonomy, while static analysis verifies interface registrations and executable effect sites. This stage produces a Functionality List that records an interface anchor and an effect candidate for each security-sensitive functionality (Section~\ref{subsec:high-risk}). For each entry, Bulkhead confirms the external input field as the source and the privileged operation or dependency boundary as the sink. It recovers the Relevant Call Chain by intersecting forward reachability from the source with a backward slice from the sink over the Joern Code Property Graph (CPG) \cite{joern}. The retained chain preserves the path transformations, checks, resolutions, and control predicates needed for detection (Section~\ref{subsec:recovery}). 
Finally, Bulkhead maps the summarized patterns to the function call chain so as to guide staged semantic analysis under the execution context. This process grounds abstract security relations in repository-specific evidence rather than relying on an unconstrained LLM judgment. The resulting candidates are validated through isolated PoC execution (Section~\ref{subsec:detection}).

\subsection{High Risk Functionality Discovery}
\label{subsec:high-risk}
This stage identifies functionalities in the repository that may cause a host-side component to operate on container-controlled filesystem objects. For a given repository, Bulkhead fixes the source revision at a specified commit and its corresponding dependency versions. It then derives the Target Source Set for the selected build target, platform, and build configuration. This set contains the repository source files compiled into that target and defines the scope of subsequent functionality discovery. 

\vspace{3pt}\noindent\textbf{Documentation Side Discovery.}
Documentation and source code in repository expose different parts of a functionality. Documentation describes intended behavior and public interfaces, but may be incomplete or use project-specific terms. Source code provides implementation evidence, but an external operation may be hidden behind interface registration, wrapper functions, or low-level APIs. Therefore, Bulkhead collects evidence from two aspects. The documentation side examines README files, API specifications, configuration guides, and executable examples associated with the selected build target. For each relevant passage, the LLM extracts the functionality name used by the project, its stated behavior, candidate interfaces and configuration fields. For example, mount a host directory may map to \texttt{filesystem.mount.bind}, whereas populate files into a workspace may map to \texttt{workspace.file.copy}. These mappings serve only as semantic candidates. Bulkhead requires code evidence before treating a functionality as implemented.

\vspace{3pt}\noindent\textbf{Code Side Discovery.}
Code evidence covers the entry and effect sides. The entry side determines how a container user invokes the functionality. Bulkhead summarizes functions in the Target Source Set to recover their interface roles, external inputs and wrapper relations. High-risk entry forms extracted from historical vulnerabilities guide the search for command registrations, API routes, RPC handlers, configuration callbacks, and plugin hooks. Bulkhead combines these forms with documentation-derived identifiers and function summaries. AST and type analysis verify each registration under the selected configuration, resolve its handler or callback, and identify the field or parameter carrying external input as a source candidate. The effect side identifies the filesystem or container operation reached from the entry. High-risk effects extracted from historical vulnerabilities guide the search for relevant system calls, library functions, dependency APIs, and external commands. Bulkhead locates candidate primitives with \texttt{ripgrep} and retains executable call sites using Semgrep, AST analysis, and type analysis. The LLM analyzes the function summaries and supporting code to classify the relevant calls or wrappers. The resulting executable call site becomes a sink candidate.
 
After collecting evidence, Bulkhead consolidates them into a Functionality List. The LLM interprets project-specific terminology and maps each evidence item to a common \texttt{feature\_id}. Evidence describing the same functionality and implementation is merged into one entry. For example, \texttt{request.Mounts}, and \texttt{runtime.Mount} are normalized to \texttt{filesystem.mount.bind}.

\subsection{Source-to-Sink Context Recovery}
\label{subsec:recovery}
\begin{figure}[t]
	\centering
	\includegraphics[width=.48\textwidth]{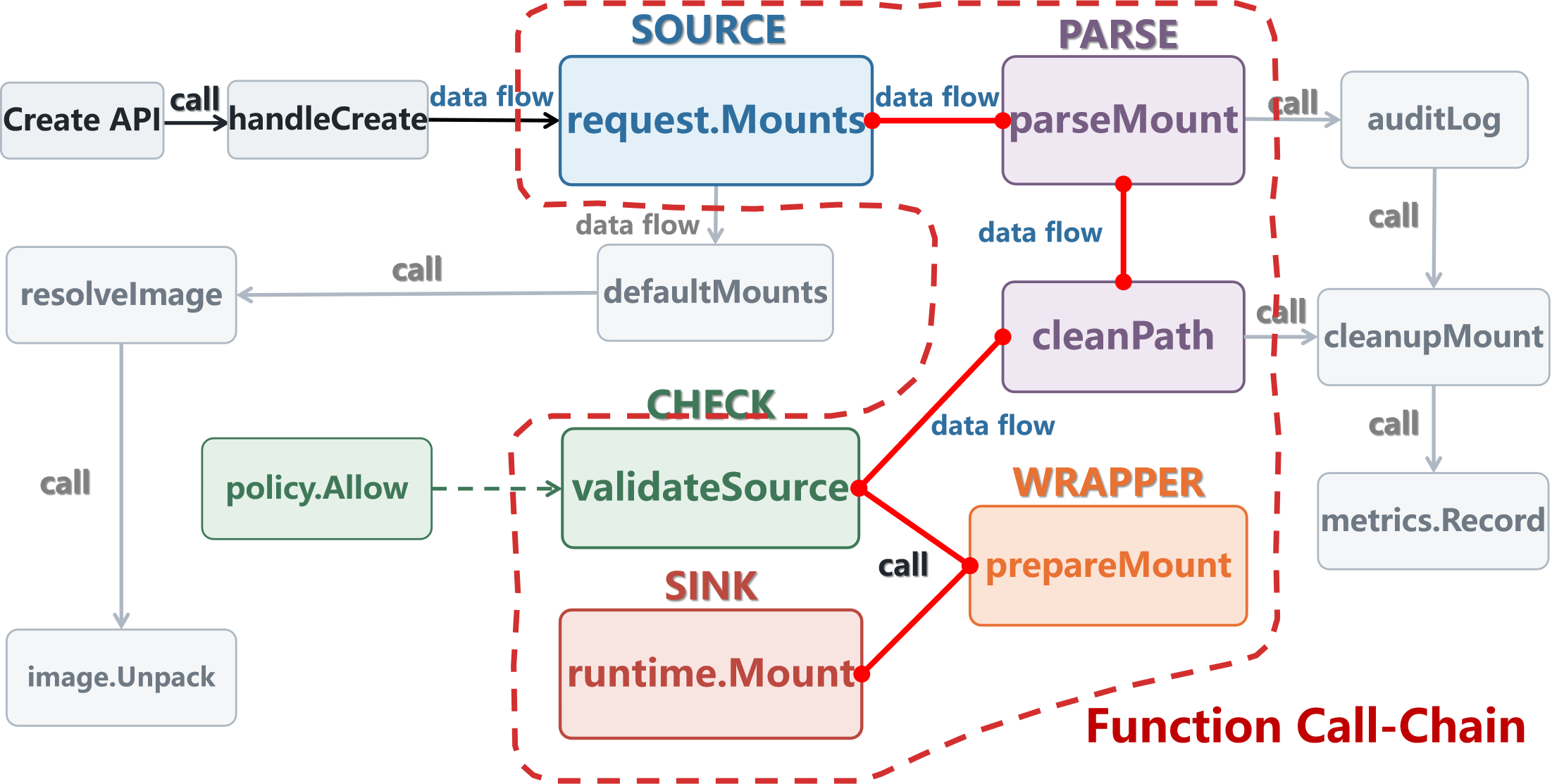}
	\caption{An Example of Source-to-Sink Context Recovery.} 
	\label{fig:call-chain}
\end{figure}

The Functionality List identifies an external interface and an effect candidate, but not the input that reaches the effect or the checks governing its use. Bulkhead recovers this missing context as a Relevant Call Chain as shown in Fig.~\ref{fig:call-chain}. The chain retains the calls, statements, data flows, and control predicates required for security analysis.

\vspace{3pt}\noindent\textbf{Source and Sink Confirmation.} 
For each Functionality List entry, Bulkhead builds a CPG over the corresponding Target Source Set using Joern. The graph first captures function calls, argument and return-value propagation, definition-use relations, and control dependencies. Bulkhead then confirms the source and sink from the primary interface and effect candidate. It follows the statically resolved registration from the interface to its handler and identifies the field or parameter carrying external input. This value is the source, while the interface serves only as an entry anchor. For example, a mount API may resolve to a creation handler, while \texttt{request.Mounts} is the actual source through which a user-controlled path enters the implementation. Bulkhead confirms the sink at an executable call site that performs the cataloged effect, such as \texttt{runtime.Mount} or \texttt{archive.Apply}. If the candidate for the effect is a project-specific wrapper, Bulkhead will track the called functions after they have been parsed until it reaches a system call, a container API, or a confirmed dependency boundary.

\vspace{3pt}\noindent\textbf{Bidirectional Chain Recovery.} 
Bulkhead performs forward data flow analysis from the source and computes a backward slice from the sink. The forward analysis follows the external value through assignments, fields, function arguments, and return values. The backward slice traces the values passed to the sink together with the callers and control conditions that determine whether the operation executes. Their intersection forms the candidate subgraph.

Unbounded interprocedural expansion may reach multiple wrappers and unrelated callees. Bulkhead therefore derives a practical depth bound from historical vulnerabilities. For each historical case, the analysis records the number of interprocedural call edges between the confirmed source and sink. The observed values define a depth range, whose upper end bounds call graph expansion during recovery. This setting covers the structural depths found in historical cases while keeping the analysis tractable. Chains beyond this range remain outside the detector's current coverage.

Bulkhead provides the candidate subgraph and its associated code context to the LLM. The LLM identifies intermediate nodes that parse, transform, resolve, validate, or prepare the path for privileged use and assigns evidence-backed roles such as parse, transform, resolve, check, and wrapper. The source and sink remain fixed by the program analysis. Bulkhead retains the labeled nodes and any propagation nodes needed to connect them through existing CPG edges. Nodes with no role in path handling or in governing sink execution are discarded. The LLM can interpret project-specific wrappers, but cannot introduce a call, data-flow, or control-dependence relation absent from the CPG. For example, a recovered chain may contain \texttt{request.Mounts}, \texttt{parseMount}, \texttt{cleanPath}, \texttt{validateSource}, \texttt{prepareMount}, and \texttt{runtime.Mount}.

The output is an ordered representation of the retained CPG subgraph, termed the Relevant Call Chain. Each node records its semantic role, qualified symbol, source location, and relevant statement, while each edge preserves the recovered call, data-flow, or control-dependence relation. The subsequent detection consumes this structured representation to determine whether the interaction violates the security invariant.

\subsection{Pattern Guided Vulnerability Detection}
\label{subsec:detection}
Bulkhead constructs vulnerability patterns from historical container path traversal cases and uses them to guide LLM detection over each Relevant Call Chain. It then validates reported candidates through isolated PoC execution.

\vspace{3pt}\noindent\textbf{Container Path Traversal Pattern Abstraction.} 
Historical cases share one failure: a privileged host component uses a path derived from a container-controlled filesystem without preserving the authorized boundary or the identity of the validated object. Bulkhead summarizes these cases along two dimensions. Lifecycle distinguishes pre-start from runtime and defines when the attacker can modify the object and trigger a host operation. Root cause distinguishes misresolution from TOCTOU. Misresolution allows privileged path resolution to escape its boundary or use an unintended filesystem view. TOCTOU allows the attacker to replace a pathname or its ancestor between validation and use. The two dimensions produce four patterns covering both lifecycle and root causes.

Each pattern contains object, interaction, and exploitation. Object records the protected object, authorized boundary, participant privileges, attacker control, and filesystem views. Interaction records the ordered checks, resolutions, modifications, and uses, together with the principal performing each operation. Exploitation records the attacker actions and state changes required to violate the invariant. The representation omits project-specific APIs while retaining the relations needed for matching and validation.


\begin{figure*}[t]
  \centering
  \includegraphics[width=0.98\textwidth]{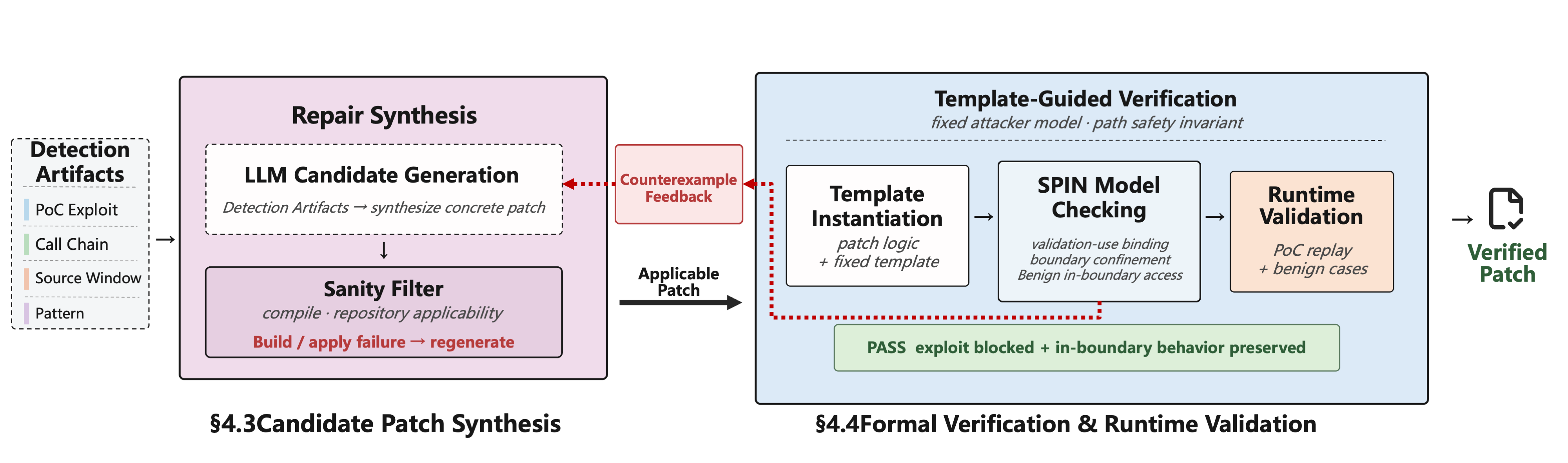}
  \caption{Bulkhead Remediation Pipeline.}
  \label{fig:patch_pipeline}
\end{figure*}

\vspace{3pt}\noindent\textbf{Pattern Guided LLM Detection and Validation.} 
Bulkhead analyzes each Relevant Call Chain with a three-stage derived from the pattern structure. The object stage binds pattern roles to concrete path variables, principals, boundaries, and filesystem views found at the confirmed source and sink. These bindings become fixed inputs to the interaction stage, which maps each required check, resolution, modification, and use to a chain node. This stage verifies execution order, the principal performing each operation, and whether validation and use refer to the same filesystem object. The exploitation stage checks whether the recovered interface and program state provide the mutation opportunity and trigger required by the pattern. Each assignment must cite a node, edge, or source location. Static analysis fixes the source, sink, call relations, data flow, and control dependencies. The LLM resolves semantic roles and project-specific wrappers but cannot alter these facts or introduce missing relations. Bulkhead reports a candidate only when the required pattern relations are jointly supported by the chain.

The detection result records the matched pattern, problem details, and threat model. Bulkhead also instantiates the mapped exploitation field as a PoC and executes it in isolation. Only a demonstrated boundary violation is marked as validated. A formalized bug report summarizes the violated invariant, supporting code, exploitation conditions, and validation result for remediation.

%% file: 4-patch.tex
\section{LLM-based Logic Flaw Remediation}
\label{sec:patch_pipeline}

\subsection{Challenges of Remediating Logic Flaws}
\label{subsec:patch_challenges}

{\color{black}
Remediating a logic flaw poses two problems beyond detecting it. A repair must first restore the violated validation-to-use relation without removing legitimate behavior. Its verification must then preserve the candidate's security-relevant behavior in a model whose attacker semantics and properties are not chosen by that candidate.

\vspace{3pt}\noindent\textbf{Challenge I: Establishing Repair Adequacy beyond Exploit Suppression.}
A failed PoC does not establish an adequate repair. A PoC covers one input and execution path, so a candidate can stop it by recognizing a specific value, rejecting the affected operation, or moving the unsafe behavior to an untested path. These changes suppress the observed exploit without necessarily restoring the violated relation.

Pouch~1.3.1 illustrates how conservative symlink handling can break legitimate behavior. Its copy-out path leaves symlink following unsupported and calls \texttt{CopyInfoSourcePath} with \texttt{followLink=false}, so a final symlink is archived as a link rather than followed even when its target remains inside the container root filesystem~\cite{aliyun2022pouchcp}. This policy sacrifices legitimate final component dereferencing.

Conversely, adding a relevant check does not ensure that the privileged operation uses the checked object. CVE-2017-1002101 in Kubernetes \texttt{subPath} was repaired by validating path components and passing an opened \texttt{/proc/<pid>/fd/<fd>} reference to the mount operation. CVE-2021-25741 later showed that the mount utility canonicalized this reference back into a mutable pathname, allowing replacement after validation. The check existed, but validation and use were no longer bound to the same object.

An adequate repair must therefore satisfy three requirements: confine the privileged effect, preserve validation-to-use binding, and retain legitimate in-boundary layouts. PoC and benign-workload replay provide useful concrete evidence, but only for the exercised executions. They cannot establish the absence of residual vulnerabilities.

\vspace{3pt}\noindent\textbf{Challenge II: Establishing Trustworthy Source-to-Model Correspondence.}
Model checking is useful only when the verification model faithfully preserves the candidate's security-relevant behavior. A direct LLM verdict cannot be independently executed, while an unconstrained LLM-generated model may omit an attacker transition, bind validation and use to different source objects, or encode the required property as an assumption. Spin can exhaustively explore such a model and still return no counterexample.

The central challenge is faithful abstraction, not executable Promela generation. Candidate-independent semantics, including attacker actions and security and compatibility properties, must remain fixed. The LLM may construct only candidate-specific behavior, which must be bound to operations in the patched source and rejected when incomplete or vacuous. This separation makes the model inspectable and the verifier deterministic, while keeping any no-counterexample result explicitly bounded by the chosen abstraction.
}

\subsection{Remediation Pipeline}
\label{subsec:patch_overview}

{\color{black}
Bulkhead separates patch synthesis from patch acceptance. Given a validated vulnerability report, the patch pipeline receives localized source evidence and the violated validation-to-use relation, then returns a repository-specific Git unified diff without access to a human reference patch. A separate runner applies and builds the candidate, instantiates the matched verification template, and invokes Spin. When reconstruction is possible, a surviving candidate is also replayed against the original PoC and predeclared benign workloads. The LLM therefore proposes repairs, while repository checks, model checking, and runtime validation determine the evidence that each repair supports.
}

{\color{black}
A \emph{Patch Manifest} fixes the repair contract before synthesis. It records the target repository and revision, build target, entry-to-effect call chain, matched vulnerability pattern, attacker capabilities, required security property, PoC harness, and benign behaviors to preserve. The human reference patch is withheld from synthesis and used only during evaluation. Freezing this contract prevents the repair loop from redefining the vulnerability or weakening its acceptance criteria after seeing a candidate.
}

{\color{black}
Bulkhead reports each validation gate separately because gates support different claims. As shown in Fig.~\ref{fig:patch_pipeline}, repository checks establish only that a candidate applies to the pinned revision and builds at the affected target. Pattern-template model checking evaluates the required security property within the selected model abstraction and attacker model. Runtime replay, when available, tests the original exploit and the benign behaviors covered by the harness. A \emph{Patch Artifact} records the candidate and the evidence from every completed gate. Missing runtime evidence remains explicit, and a model-checking result is not presented as unrestricted source-level correctness.
}

\subsection{LLM-based Patch Synthesis}
\label{subsec:patch_synthesis}

{\color{black}
Patch synthesis is restricted to a frozen, evidence-bounded input. The SDK-based provider receives the PoC attack summary, the Joern-derived entry-to-effect call chain, line-anchored source windows from the pinned vulnerable revision, and the security constraint defined by the matched pattern. It receives neither the repository checkout nor the human reference patch, sealed confirmation profile, or hidden harness details, and it runs without tool or workspace access. These restrictions expose the repair-relevant context without allowing the provider to inspect acceptance artifacts.
}

{\color{black}
The synthesis boundary is enforced through a single output format and a deterministic repository gate. The provider must return one complete Git-format unified diff; function fragments, prose, Markdown fences, and edits outside the allowed paths are rejected, and every raw response is retained. A runner-owned canonicalizer applies the response to a clean checkout and may normalize only transport defects that preserve the resulting source tree, such as a missing final newline or inconsistent hunk counts accepted by Git recounting. A separate evaluator reapplies the canonical diff to a fresh checkout, checks whitespace and changed-file scope, and runs the case-specific build. Passing this gate establishes only that the candidate applies to the repository; it does not show that the candidate satisfies the security invariant.
}

{\color{black}
Feedback is typed and bounded so that iteration remains auditable. Each trial admits one initial candidate and at most two feedback-guided repairs. Applicability and build failures return \texttt{PATCH\_APPLY\_FAILED} or \texttt{BUILD\_FAILED} with a bounded diagnostic excerpt; model checking may return \texttt{MODEL\_COUNTEREXAMPLE}. Every attempt is retained, and the trial stops when a candidate clears the enabled checks, encounters a terminal error, or exhausts the three-candidate budget. Only repository-applicable candidates enter model checking.
}

\subsection{Pattern-Template Model Checking}
\label{subsec:template_mc}

{\color{black}
Pattern-template model checking evaluates the frozen repair contract rather than repository applicability. The matched vulnerability pattern selects a hash-bound Promela template that fixes the attacker process, finite path-state domain, relevant interleavings, and required safety property. When legitimate in-boundary behavior must be preserved, the same template also fixes compatibility assertions. These candidate-independent elements remain unchanged across a trial.
}

{\color{black}
Candidate-specific abstraction is confined to designated template slots. From the canonical diff and the frozen verification context, an LLM-assisted translator supplies statement-level Promela snippets only for declared placeholders. Each snippet must bind to a distinct added source line and to predefined model predicates, while initialization remains template-owned. Before rendering, the backend rejects missing or additional fields, empty or vacuous snippets, structural Promela constructs, unbound source evidence, and assignments that directly force a frozen safety predicate to hold. These checks establish a well-formed, source-bound model instance, not semantic equivalence with the source. Bulkhead therefore retains the candidate, template, mapping, generated model, and their cryptographic digests for inspection.
}

{\color{black}
Verifier outcomes are interpreted conservatively. Spin~\cite{holzmann1997model} reports \texttt{safe} only after a complete run with zero property violations and \texttt{unsafe} when it finds a concrete violation. Syntax or compilation failures, timeouts, infrastructure failures, and incomplete searches produce \texttt{error}; a repository gate that prevents verification produces \texttt{not-run}. None of these latter outcomes is treated as safety evidence.
}

{\color{black}
Counterexamples guide repair without overstating what the model proves. For an \texttt{unsafe} result, Bulkhead retains the trace and returns a bounded excerpt as \texttt{MODEL\_COUNTEREXAMPLE}, subject to the shared three-candidate budget. Depending on the template, the trace may expose an unbound validation-to-use sequence, an unchecked out-of-boundary target, or a regression in legitimate in-boundary behavior. It demonstrates a violating execution in the instantiated model, not source-level exploitability by itself. A \texttt{safe} candidate proceeds to runtime replay when reconstruction is possible, and the Patch Artifact keeps repository, bounded model, and runtime evidence as separate outcomes.
}


%% file: 5-evaluation.tex
\section{Evaluation}
\label{sec:evaluation}

\subsection{Overview}
\label{subsec:eval-setup}

\vspace{3pt}\noindent\textbf{Research Questions.}
Our evaluation aims to answer the following questions:
\begin{itemize}
    \item How effective is Bulkhead at discovering unknown vulnerabilities in real-world container projects?
    \item How effectively does Bulkhead detect vulnerabilities on a leakage-controlled dataset?
    \item {\color{black}How effectively do patches generated by Bulkhead prevent exploits while preserving intended functionality?}
    \item How do core components affect the detection and remediation performance of Bulkhead?
\end{itemize}

\vspace{3pt}\noindent\textbf{Evaluation Units.}
We use repositories as units for real-world analysis and functionalities for effectiveness evaluation. In the real-world study (Section~\ref{subsec:discovery}), each repository snapshot pinned to a commit is processed end to end for detection and remediation. In the effectiveness evaluation (Section~\ref{subsec:Effectiveness Detection}, Section~\ref{Effectiveness Remediation}) and ablation study (Section~\ref{subsec:ablation}), each independently labeled host--container interaction functionality in a fixed repository version forms one detection unit, and each vulnerable unit defines one remediation task using Bulkhead's validated bug report. {\color{black}For the controlled detection study, the dataset contains 25 CVEs disclosed after September 30, 2024, the knowledge cutoff of GPT-5.1. The remediation study reuses the same vulnerable units, while Bulkhead's vulnerability patterns are derived only from cases disclosed before September 30, 2024.}

\subsection{Real-World Vulnerability Discovery}
\label{subsec:discovery}

\vspace{3pt}\noindent\textbf{Repository Collection.}
In order to discover path traversal vulnerabilities in real-world container projects, we systematically collected candidate repositories through GitHub topic keyword searches, relevant links within the main container ecosystem, and public repositories maintained by container-related organizations. After excluding repositories used as demos or those containing only documentation, we finally selected 82 open-source repositories for testing. These repositories cover various types of projects, including container runtimes such as Docker and Nerdctl, hardware support systems such as NVIDIA Container Toolkit and Intel Device Plugin, and container-based AI sandbox environments such as Openhands. We conducted manual audits on a stratified sample of 20\% of the total repositories to confirm that all audited repositories met the inclusion criteria.

\begin{table}[t]
    \centering
    \caption{Real-world Vulnerability Detection Results.}
    \label{tab:real-world-discovery}
    \footnotesize
    \begin{tabularx}{\linewidth}{
        @{}>{\raggedright\arraybackslash}Xcccc@{}}
        \toprule
        \textbf{Project Category} &
        \textbf{Repo.} &
        \textbf{Func.} &
        \textbf{Cand.} &
        \textbf{Conf.} \\
        \midrule
        Container Management Tools
            & 38 & 130 & 10 & 8 \\
        Container Toolkits and Build Tools
            & 31 & 122 & 18 & 11 \\
        Container-based AI Sandboxes
            & 13 & 19 & 7 & 1 \\
        \midrule
        \textbf{Total}
            & \textbf{82} & \textbf{271}
            & \textbf{35} & \textbf{20} \\
        \bottomrule
    \end{tabularx}
\end{table}

\vspace{3pt}\noindent\textbf{Detection Results.}
We run Bulkhead on 82 container-related repositories with GPT-5.6-Sol \cite{gpt5.6sol}. As summarized in Table~\ref{tab:real-world-discovery}, Bulkhead first identifies 271 host-container interaction functionalities and detect 35 of them as potential vulnerability candidates. We attempt isolated PoC validation for every candidate functionality and finally confirm 20 zero-day vulnerabilities. We have reported each confirmed vulnerability to the corresponding maintainers, together with the affected revision, root cause analysis, executable PoC, and remediation suggestions.

\begin{framed}
    \noindent\textbf{Insight I: Severity.}
    Container path traversal vulnerabilities remain a recurring security weakness in current container ecosystems.
\end{framed}

At the time of writing, 3 of the discovered vulnerabilities have been assigned CVE identifiers, while the remaining reports are undergoing maintainer review or patch development. The confirmed 20 cases occur in 18 independent repositories, showing that unsafe host-container path handling is not limited to a single implementation. Although the affected functionalities use different interfaces and filesystem operations, they can still fail to preserve the intended filesystem boundary when privileged host components process container-controlled paths. These findings show that container path traversal remains an active security problem in real-world container ecosystems.

\vspace{3pt}\noindent\textbf{Case Study.}
We present two representative cases to illustrate the diversity and impact of the discovered vulnerabilities. The first case involves Nerdctl, which represents a widely used container engine. Bulkhead identifies a pre-start vulnerability in the volume mounting procedure. The issue resides in the ProcessFlagV function in mountutil.go, which resolves user-specified mount paths during container initialization. The function validates path existence but does not enforce path stability across resolution and use. In a scenario where multiple containers share a common volume, an attacker can manipulate the shared directory from a running container and replace a legitimate subpath with a symlink that targets a sensitive host location. When a new container is started with a bind mount that references this subpath, the resolution process follows the manipulated link and mounts the host path into the container. The attack enables unauthorized access to sensitive files on the host filesystem and leads to container escape. The vulnerability reflects a violation of the invariant that path validation must remain consistent with path usage under concurrent modifications.

\begin{framed}
    \noindent \textbf{Insight II:} 
    Generalizability: Path traversal vulnerability extends beyond traditional container tools to emerging implementations, such as container-based AI sandbox.
\end{framed}

The second case involves OpenHands, which represents a container-based AI sandbox environment that relies on container isolation to execute untrusted workloads. Bulkhead identifies a runtime vulnerability in the file transfer mechanism between the client and the sandboxed Action Execution Server. In the V0 runtime, the vulnerability arises from insufficient path boundary enforcement in the copy\_from APIs. The server processes file operations using standard library functions that follow symlinks and do not restrict extraction paths to the intended workspace. An attacker with execution capability inside the sandbox can create a symlink that points to a sensitive directory outside the workspace. When the client invokes the download interface, the server traverses the symlink and packages the target files into an archive, which is then returned to the client. The attacker can extract the archive and access sensitive data. This vulnerability demonstrates that path traversal issues are not limited to traditional operations, and they also affect higher-level file interaction APIs in sandbox environments.

\subsection{Effectiveness of Detection Pipeline}
\label{subsec:Effectiveness Detection}

We construct a dedicated dataset to evaluate the detection pipeline under controlled ground truth. {\color{black}All LLM-based stages in this detection experiment use GPT-5.1 \cite{gpt5.1}, whose official knowledge cutoff is Sep. 30, 2024. We therefore select only vulnerabilities disclosed after this date to build the dataset.}

\vspace{3pt}\noindent\textbf{Dataset Construction.}
We search the National Vulnerability Database, GitHub Security Advisories, and vendor advisories for container path traversal CVEs disclosed after the GPT-5.1 knowledge cutoff. After removing duplicates and cases without a reproducible vulnerable revision, we retain 25 CVEs, each paired with a reproducible vulnerable repository revision. For each pinned revision, we manually label all host-container interaction functionalities. The affected functionality is labeled positive, while the remaining functionalities are labeled negative after manual inspection. Each functionality in one pinned revision forms a dataset unit. The resulting dataset contains 68 functionality units, comprising 25 positive and 43 negative units. During detection, Bulkhead does not receive the CVE descriptions, reference patches, or ground-truth labels.

\begin{table}[t]
  \centering
  \setlength{\belowcaptionskip}{2.5pt}
  \caption{{\color{black}Detection and remediation effectiveness.}}
  \label{tab:detection-effectiveness}
  \small
  \setlength{\tabcolsep}{3pt}
  \renewcommand{\arraystretch}{1.08}
  \begin{tabular*}{\linewidth}{@{\extracolsep{\fill}}lccccc@{}}
    \toprule
    \multicolumn{6}{l}{\textit{(a) Detection Pipeline Comparison.}} \\
    \midrule
    \textbf{Method}
    & \textbf{TP}
    & \textbf{FP}
    & \textbf{FN}
    & \textbf{Pre.}
    & \textbf{Rec.} \\
    \midrule
    Bulkhead + GPT-5.1
        & 23 & 5 & 2 & 82.1\% & 92.0\% \\
    Antaeus \cite{armillotta2026antaeus} + GPT-5.1
        & 8 & 12 & 17 & 40.0\% & 32.0\% \\
    Codex + GPT-5.6-Sol
        & 13 & 9 & 12 & 59.1\% & 52.0\% \\
    Claude Code + GLM-5.3
        & 11 & 8 & 14 & 57.9\% & 44.0\% \\
    \bottomrule
  \end{tabular*}

  \vspace{5pt}

  {\color{black}
  \begin{tabularx}{\linewidth}{@{}Xccc@{}}
    \toprule
    \multicolumn{4}{l}{\textit{(b) Remediation comparison}} \\
    \midrule
    \textbf{System}
    & \shortstack{\textbf{Build} $\uparrow$}
    & \shortstack{\textbf{Exploit} $\downarrow$}
    & \shortstack{\textbf{Validated} $\uparrow$} \\
    \midrule
    Bulkhead + GPT-5.1
      & \shortstack{96.0\%\\[-1pt]{\scriptsize(24/25)}}
      & \shortstack{0.0\%\\[-1pt]{\scriptsize(0/24)}}
      & \shortstack{96.0\%\\[-1pt]{\scriptsize(24/25)}} \\
    OpenHands~\cite{wang2025openhands} + GPT-5.1
      & \shortstack{72.0\%\\[-1pt]{\scriptsize(18/25)}}
      & \shortstack{33.3\%\\[-1pt]{\scriptsize(6/18)}}
      & \shortstack{48.0\%\\[-1pt]{\scriptsize(12/25)}} \\
    SWE-agent~\cite{yang2024sweagent} + GPT-5.1
      & \shortstack{72.0\%\\[-1pt]{\scriptsize(18/25)}}
      & \shortstack{33.3\%\\[-1pt]{\scriptsize(6/18)}}
      & \shortstack{48.0\%\\[-1pt]{\scriptsize(12/25)}} \\
    Codex + GPT-5.6-Sol 
      & \shortstack{88.0\%\\[-1pt]{\scriptsize(22/25)}}
      & \shortstack{18.2\%\\[-1pt]{\scriptsize(4/22)}}
      & \shortstack{72.0\%\\[-1pt]{\scriptsize(18/25)}} \\
    \bottomrule
  \end{tabularx}

  \vspace{1pt}
  \parbox{\linewidth}{\scriptsize
  Panel~(b) reports Build and Validated rates over all 25 tasks.
  Exploit success is the fraction of buildable patches for which the original exploit still succeeds or affect the normal function.
  A Validated patch must apply and build, pass model checking, block the exploit, and preserve benign behavior.}
  }
  \vspace{2pt}
\end{table}

\vspace{3pt}\noindent\textbf{Results of Detection Effectiveness.}
Table~\ref{tab:detection-effectiveness} panel~(a) compares Bulkhead with the baseline Antaeus \cite{armillotta2026antaeus} on GPT-5.1 and two contemporary repository-level coding agents on the same dataset. Bulkhead and Antaeus invoke GPT-5.1 through OpenRouter. GPT-5.1 has a knowledge cutoff in 2024, which predates the public disclosure of the dataset. Codex with GPT-5.6-Sol and Claude Code with GLM-5.3 represent contemporary repository-level coding agents \cite{glm}. We provide both agents with the same repository snapshots and a common prompt describing the detection target.

Bulkhead achieves the highest precision and recall while producing substantially fewer false positives and false negatives than Antaeus. Among the two coding agents, Codex performs slightly better than Claude Code, but both remain well below Bulkhead in recall and produce more false positives. These results show that Bulkhead's advantage does not arise solely from LLM capability: the controlled GPT-5.1 comparison isolates benefits of its workflow, while the coding-agent comparison shows that prompt-guided repository inspection does not reliably recover security relation between container-controlled paths and privileged host operations.

\subsection{Effectiveness of Remediation Pipeline}
\label{Effectiveness Remediation}

{\color{black}
We evaluate remediation in two settings. The real-world study asks whether Bulkhead can turn confirmed findings into validated repairs. The controlled study uses the same 25 historical CVE cases as Section~\ref{subsec:Effectiveness Detection} to compare repair configurations under identical task inputs and candidate budgets.

\vspace{3pt}\noindent\textbf{Remediation of Newly Discovered Vulnerabilities.}
We first evaluate whether Bulkhead can turn a confirmed vulnerability report into a repair that passes both repository and security checks.
Each of the 20 findings reported in our real-world study defines a separate repair task.
Bulkhead employs GPT-5.6-Sol to generate patches.
Its input follows the frozen repair contract and contains the vulnerable revision, the relevant call chain and source window, the matched vulnerability pattern, and the required security property.
The LLM returns one complete Git unified diff.
Bulkhead applies the candidate to a clean checkout and runs build target specified for that case.

A buildable patch is not necessarily valid. Bulkhead therefore performs pattern-template model checking and runtime replay on both vulnerable and patched revisions, including benign-path tests. A repair is fully validated only if it passes all repository, model, exploit, and compatibility checks; missing or failed gates are reported separately. These claims are limited to the modeled properties and executed oracles. We manually reviewed all 20 cases and model artifacts for consistency with the underlying vulnerability semantics. We then submitted our patch suggestions to the corresponding upstream maintainers.

\vspace{3pt}\noindent\textbf{Dataset Construction.}
The controlled remediation experiment reuses the 25 container path traversal CVEs disclosed after September 30, 2024, as described in Section~\ref{subsec:Effectiveness Detection}.
To isolate remediation performance, we use human-validated vulnerability reports for all cases. Thus the reported result measures remediation performance given a correct report, rather than end-to-end system performance.
For each case, we define two runtime oracles before patch synthesis. The exploit oracle first confirms that the original PoC succeeds on the vulnerable revision and then requires it to be blocked on the patched revision. The functionality oracle executes a predeclared set of normal commands and checks their expected exit status and observable behavior. A patch passes dynamic validation only when the exploit oracle confirms that the PoC is blocked and all normal-command tests pass.
End-to-end validity requires the candidate to pass the repository checks, satisfy the encoded model properties, and pass dynamic validation. We report a missing or failed gate separately and do not count that candidate as an end-to-end valid repair.

\vspace{3pt}\noindent\textbf{Results of Remediation Effectiveness.}
Table~\ref{tab:detection-effectiveness}, panel~(b), reports system-level outcomes on the 25 remediation tasks. Bulkhead produces 24 fully validated repairs (96.0\%), compared with 18 (72.0\%) for Codex + GPT-5.6-Sol in the prompt-only setting and 12 (48.0\%) each for OpenHands and SWE-agent. It also achieves the highest repository-level success: 24 of 25 patches apply and build, versus 22 for Codex and 18 for each of the other two systems.

The security oracle further separates buildable patches from effective repairs. None of Bulkhead's 24 buildable patches leaves the exploit effective, whereas the exploit still succeeds against 4 of Codex's 22 buildable patches and 6 of 18 for both OpenHands and SWE-agent. Because the Exploit column is conditioned on buildable patches while Validated uses all 25 tasks, the end-to-end comparison is given by the Validated column. A patch is counted there only after it also passes model checking, blocks the exploit, and preserves benign behavior. Under the reported configurations, Bulkhead's validated repair rate therefore exceeds prompt-only Codex by 24 percentage points and each repository-level agent by 48 percentage points.

\vspace{3pt}\noindent\textbf{Compatibility Analysis.}
Exploit rejection alone is an incomplete repair criterion.
Three historical fixes stop the reported boundary escape by rejecting symbolic links entirely.
This response also rejects legitimate symbolic links whose resolved targets remain inside the intended filesystem boundary.
Bulkhead represents the required benign behavior as a compatibility property and rejects candidates that violate it.

\begin{framed}
    \noindent \textbf{Insight III: Compatibility.}
    Historical fixes confirm that blocking the reported exploit does not establish a valid repair. Bulkhead checks the corresponding compatibility property and rejects candidates that disable legitimate symbolic links inside the intended filesystem boundary.
\end{framed}

Security property excludes paths that escape the intended boundary, whereas the compatibility property preserves valid path layouts inside it.
A repair must satisfy both properties to avoid replacing a security flaw with a functional regression.
}

\subsection{Ablation Study}
\label{subsec:ablation}

{\color{black}We evaluate detection and remediation separately using the same datasets and LLM configurations as Sections~\ref{subsec:Effectiveness Detection} and~\ref{Effectiveness Remediation}. The detection study removes one component from the complete pipeline at a time. The remediation study compares the full pipeline with a matched variant that withholds model-checking feedback from synthesis. Both configurations use the same repair context, synthesis LLM, candidate budget, repository diagnostics, and final validation gates; only the counterexample feedback returned after a model-property violation differs.}

\begin{table}[t]
\centering
\setlength{\belowcaptionskip}{2.5pt}
\caption{{\color{black}Ablation Results for Detection and Remediation.}}
\label{tab:ablation}
\footnotesize
\setlength{\tabcolsep}{3pt}

{\color{black}
\begin{tabularx}{\columnwidth}{@{}Xccccc@{}}
\toprule
\multicolumn{6}{l}{\textit{(a) Detection pipeline}} \\
\textbf{Configuration}
& \textbf{TP}
& \textbf{FP}
& \textbf{FN}
& \shortstack{\textbf{Precision}}
& \shortstack{\textbf{Recall}} \\
\midrule
Full
& 23 & 5  & 2  & 82.1\% & 92.0\% \\
w/o Functionality Discovery
& 19 & 15 & 6  & 55.9\% & 76.0\% \\
w/o Context Recovery
& 13 & 12 & 12 & 52.0\% & 52.0\% \\
w/o Pattern Guidance
& 8  & 11 & 17 & 42.1\% & 32.0\% \\
\bottomrule
\end{tabularx}

\vspace{3pt}
{
\setlength{\tabcolsep}{2.2pt}
\begin{tabularx}{\columnwidth}{@{}Xccc@{}}
\toprule
\multicolumn{4}{l}{\textit{(b) Remediation pipeline}} \\
\textbf{Configuration}
& \shortstack{\textbf{Build}$\uparrow$}
& \shortstack{\textbf{Exploit} $\downarrow$}
& \shortstack{\textbf{Validated} $\uparrow$} \\
\midrule
\shortstack[l]{w/o Model-Checking Feedback}
& \shortstack{96.0\%\\(24/25)}
& \shortstack{29.2\%\\(7/24)}
& \shortstack{68.0\%\\(17/25)} \\
Full
& \shortstack{96.0\%\\(24/25)}
& \shortstack{0.0\%\\(0/24)}
& \shortstack{96.0\%\\(24/25)} \\
\bottomrule
\end{tabularx}

\vspace{1pt}
\parbox{\columnwidth}{\scriptsize
Panel~(b) reports Build and Validated rates over all 25 tasks.
  Exploit success is the fraction of buildable patches for which the original exploit still succeeds or affect the normal function.
  A Validated patch must apply and build, pass model checking, block the exploit, and preserve benign behavior.}
}
}
\end{table}



\vspace{3pt}\noindent\textbf{Detection Pipeline.}
Table~\ref{tab:ablation} reveals three distinct failure modes in the detection pipeline. The complete pipeline detects 23 vulnerabilities with 5 false positives and 2 false negatives, achieving 82.1\% precision and 92.0\% recall. High Risk Functionality Discovery determines which deployment-related host--container interactions enter the analysis and supplies the project-specific semantics needed to interpret them. Removing this stage reduces true positives from 23 to 19, increases false positives from 5 to 15, and lowers precision to 55.9\%. Documentation evidence here is particularly important for resolving interface semantics. For example, the \texttt{nerdctl cp} documentation restricts the operation to trusted and cooperative containers and does not claim protection against attacker-controlled containers~\cite{nerdctlcp}. This evidence exposes a trust assumption that cannot be inferred reliably from filesystem operations alone.

Source-to-Sink Context Recovery preserves the propagation paths and control dependencies connecting container-controlled inputs to privileged effects. Without this stage, true positives fall to 13, while false positives and false negatives increase to 12 each. The resulting precision and recall are both 52.0\%, showing that incomplete or irrelevant context causes both missed data flows and unsupported associations. Removing Invariant Guided Vulnerability Detection produces the largest loss. The pipeline detects only 8 vulnerabilities and misses 17, reducing precision to 42.1\% and recall to 32.0\%. The result identifies invariant guidance as the principal source of detection accuracy. Without an explicit pattern, the LLM does not consistently preserve the relations among attacker control, filesystem views and execution order.

\vspace{3pt}\noindent\textbf{Remediation Pipeline.}
{\color{black}
The remediation ablation reuses the 25 CVE repair tasks from Section~\ref{Effectiveness Remediation}. Table~\ref{tab:ablation}, panel~(b), compares Full with a matched variant that withholds model-checking feedback from synthesis. Both configurations retain the same candidate budget and repository diagnostics. Model checking remains part of the evaluator in both configurations, but only Full returns a bounded counterexample to the synthesis LLM after a property violation. Exploit and benign-behavior outcomes remain sealed in both configurations and are used only to determine fully validated repairs.

We use model checking because a free-form LLM judgment provides no independently checkable derivation. Instead, Bulkhead asks the LLM to instantiate an executable abstraction and obtains the verdict from SPIN's state-space exploration. This is analogous to asking the LLM to construct a program whose execution answers the question rather than accepting its answer directly: the checker produces a reproducible verdict and, on failure, a concrete counterexample.

Unconstrained model generation is insufficient, however. Our manual inspection found that freely generated Promela could omit or distort the source program's control flow and validation-to-use relations. Bulkhead therefore uses human-authored, pattern-specific templates that fix the verifier skeleton, admissible attacker actions, and property schema; the LLM fills only source-bound slots. This division preserves expert-specified semantics while automating case-specific instantiation.

Accordingly, we make only a bounded formal-verification claim. A safe verdict means that SPIN exhausts the instantiated finite state space without finding a violation of the encoded properties under the selected template and attacker model. It does not establish that the patch is vulnerability-free outside this boundary. Exploit replay and benign-behavior tests therefore remain independent validation gates.

Removing model-checking feedback leaves repository applicability unchanged as the apply check is performed before this part. The difference appears after this gate. Without feedback, 7 of the 24 buildable patches violate a model property (29.2\%), and 17 of 25 tasks yield fully validated repairs (68.0\%). With feedback, Full reports no property violations among the 24 buildable patches and fully validates 24 of 25 tasks (96.0\%), a gain of seven repairs, or 28 percentage points. Because the apply/build count is identical, the observed gain occurs after repository integration and supports the utility of model-checking counterexamples for refining property-preserving repairs on these 25 tasks.
}


\begin{table}[t]
    \centering
    \caption{Time and Token Breakdown per Implementation.}
    \label{tab:time-token-breakdown}
    \begin{tabular*}{\columnwidth}{
        @{\extracolsep{\fill}} l c c
    }
        \toprule
        \textbf{Stage}
        & \textbf{Avg. Time (s)}
        & \textbf{Tokens (M)} \\
        \midrule
        Detection   & 428 & 5.8 \\
        Remediation & 200 & 2.5 \\
        \midrule
        \textbf{Total} & \textbf{628} & \textbf{8.3} \\
        \bottomrule
    \end{tabular*}
\end{table}

\vspace{3pt}\noindent\textbf{{\color{black}Cost.}}
{\color{black}
Table~\ref{tab:time-token-breakdown} reports the average wall-clock time and LLM token consumption per target implementation for each pipeline stage. Detection takes 428 seconds and consumes 5.8 million tokens on average, whereas remediation takes 200 seconds and consumes 2.5 million tokens.
}


%% file: 6-relatedwork.tex
\section{Related Work}

\vspace{3pt}\noindent\textbf{Container Isolation Security}. 
To mitigate container escapes, researchers have focused on hardening the kernel's logical boundaries. Li et al. \cite{li2023lost} identified systematic path-misresolution vulnerabilities and addressed them through Patrol, a kernel extension that enforces dentry-based access control during path lookups. Shifting from path logic to data integrity, Xu et al. \cite{xu2024condo} introduced Condo, which utilizes trusted execution environments to protect critical kernel permission structures. Unlike localized fixes, Condo provides a transparent, low-overhead scheme for safeguarding non-control flow data against unauthorized tampering. 

Beyond software-only defenses, leveraging hardware primitives has become a key trend for achieving strong isolation from untrusted hosts. Zhou et al. \cite{zhou2025rcontainer} leverage ARM CCA hardware primitives to minimize the trusted computing base. Their RContainer architecture employs per-container shims and a lightweight mini-OS to isolate workloads from untrusted hosts, effectively shifting the security boundary from kernel software to hardware-enforced realms. Wen et al. \cite{WEN2026108478} proposed a multi-level container isolation framework for cloud-edge-terminal environments, treating container processes as security subjects and writable image layers, restricted host files, and shared volumes as protected objects. The framework integrates confidentiality and integrity labels with Bell-LaPadula, Biba, and attribute-based access-control principles to prevent unauthorized cross-container and host-resource access. To adapt isolation strength under limited computing resources, the authors formulate security-policy allocation as a static attacker-defender game and employ a wolf-pack co-evolutionary algorithm to approximate its Nash equilibrium. Their simulations indicate that prioritizing stronger isolation for critical containers can improve attack resilience while maintaining acceptable computational efficiency.



\vspace{3pt}\noindent\textbf{LLM-Based Vulnerability Detection}. 
Current research focuses on enhancing the reasoning capabilities of LLMs through external knowledge integration and hybrid analysis frameworks.
Knowledge-augmented and specification-guided methods utilize historical security data to constrain LLM hallucinations and provide semantic ground truth. Du et al. \cite{du2024vul} developed a retrieval-augmented framework that distills multi-dimensional knowledge from CVEs to help LLMs distinguish between vulnerable and patched code. In domain-specific tasks, Li et al. \cite{li2025cryptoscope} combined chain-of-thought prompting with specialized cryptographic knowledge bases to verify logic compliance.

Hybrid analysis and agentic architectures integrate LLMs with traditional program analysis to address complex system interactions. Yildiz et al. \cite{yildiz2025benchmarking} constructed JITVul for pairwise repository-level evaluation and equipped ReAct agents with on-demand retrieval of callers, callees, and function definitions to distinguish vulnerable functions from their patched counterparts. Li et al. \cite{li2025everything} enriched detection prompts with execution and data-flow contexts while employing the LLM as a judge mechanism to validate the generated rationales. Li et al. \cite{li2026detecting} combined agentic planning with classic analysis primitives to achieve scalable exploration of privilege escalation flaws in microservices. Luo et al. \cite{luo2026autonomy} combined static resource-lifecycle modeling with LLM-guided directed grey-box fuzzing to uncover resource-exhaustion DoS vulnerabilities in LLM-based agents. Wei et al. \cite{wei2025advanced} introduced a multi-agent conversational framework that identifies logic vulnerabilities through iterative collaboration and a specialized buffer-of-thought mechanism.

In addition, complementary benchmarks examine whether apparent detection gains reflect genuine vulnerability reasoning. Li et al. \cite{Li2025SV} introduced SV-TrustEval-C to disentangle structural and semantic reasoning under data- and control-flow complexity and semantic perturbations, showing across eleven LLMs that vulnerability judgments still rely heavily on pattern matching.

\vspace{3pt}\noindent\textbf{LLM-Based Patch Generation}. 
Researchers utilize specialized prompting strategies and iterative feedback loops to enhance the intrinsic reasoning capabilities of LLMs for program repair. Ahmed et al. \cite{10298561} employed self-consistency techniques that identify optimal fixes by selecting the most frequent solutions across multiple reasoning paths. Kulsum \cite{kulsum2024case} integrated chain-of-thought reasoning with execution feedback from compilers and security sanitizers to refine candidate patches iteratively. Nong et al. \cite{nong2025appatch} developed an adaptive framework that extracts vulnerability semantics and employs multi-faceted validation to improve repair quality in real-world scenarios. Kim et al. \cite{kim2025logs} utilized tree-of-thought prompting to decompose the patching process into distinct stages that span from fault localization to fix strategy formulation.

Hybrid frameworks combine the generative flexibility of LLMs with the rigor of formal methods to provide stronger security guarantees. Tihanyi et al. \cite{tihanyi2025new} integrated bounded model checking with LLMs to guide the repair process through counterexamples and stack traces. This approach ensures that the model checker verifies each repaired program before final output. Orvalho et al. \cite{orvalho2025counterexample} implemented a counterexample-guided inductive synthesis loop that utilizes MaxSAT-based fault localization to generate program sketches. These sketches provide a structured template that guides LLMs toward synthesizing correct code while the system refines results through continuous test suite validation.

%% file: 7-conclusion.tex
\section{Conclusion}


In this paper, we presented an automated pipeline to detect and remediate container path traversal logic flaws. Bulkhead formalizes historical CVEs into semantic patterns and correlates documentation with complete execution contexts to ground LLM reasoning across heterogeneous container trust boundaries. Evaluating 82 mainstream repositories, Bulkhead uncovered 20 zero day vulnerabilities, representing 43\% of the publicly reported baseline over the past decade, and generated verified patches for upstream maintainers. 
Bulkhead demonstrates that extracting semantic patterns and supplying complete execution context are critical for LLMs detection.

%% file: 8-appendix.tex
\section{Bulkhead Case Study}

\begin{figure}[h]
	\centering
	\includegraphics[width=.48\textwidth]{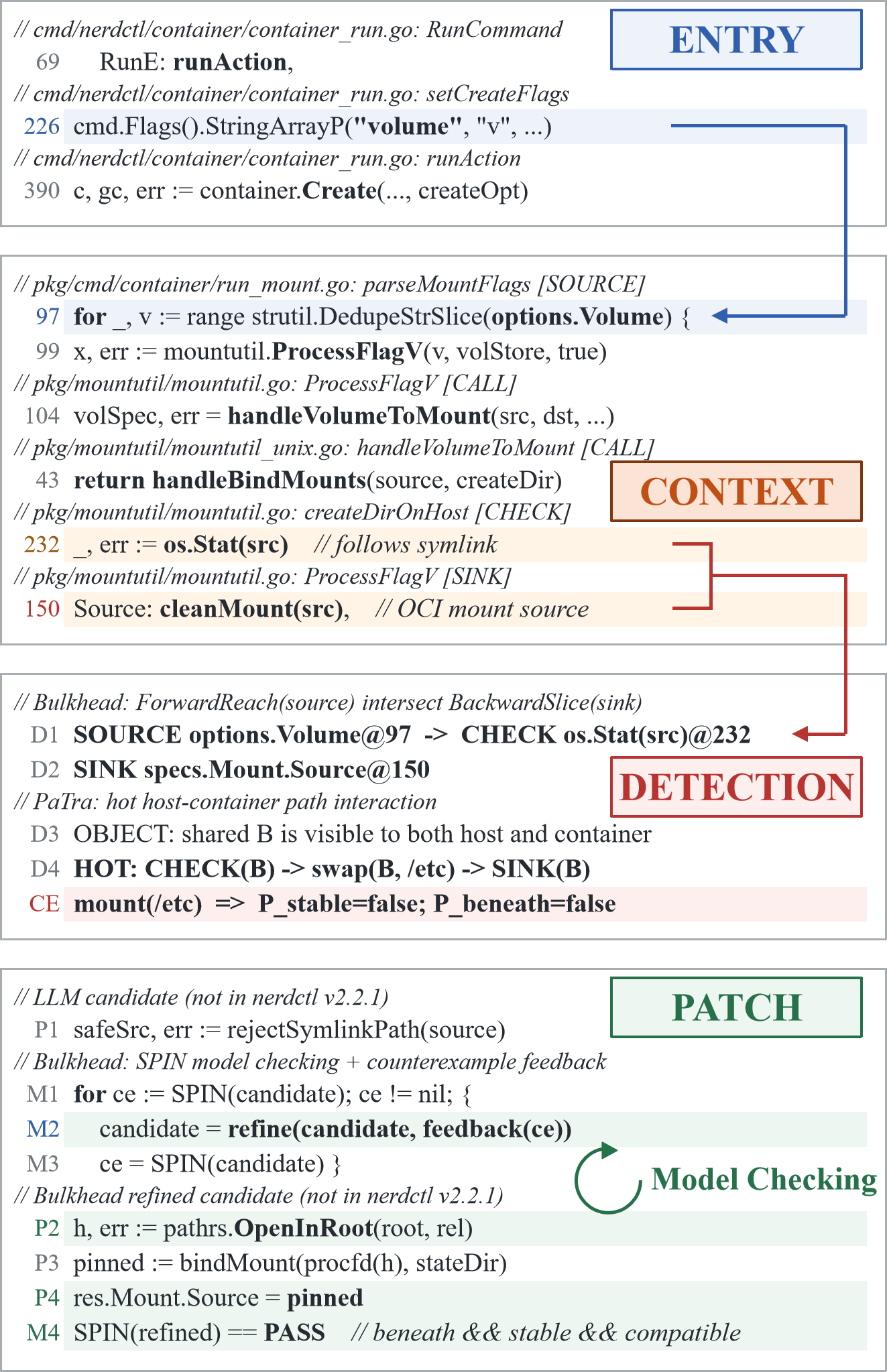}
	\caption{Bulkhead’s Real-World Detection and Remediation Workflow.} 
	\label{fig:nerdctl_mount}
\end{figure}

As shown in Figure~\ref{fig:nerdctl_mount}, the detection stage identifies high-risk host-container entry points, recovers the execution context from a container-controlled source to a privileged sink, and determines whether the interaction violates the required filesystem boundary invariants. The remediation stage uses the LLM to generate candidate patches and validates them through SPIN-based model checking. Counterexamples guide refinement until they satisfy the required security property.

\section{Vulnerability Collection}

Table~\ref{tab:path-traversal-vulnerabilities} summarizes the container path traversal vulnerabilities we collected from the past decade (2017--2026), covering a range of container-related components and categorized by their affected lifecycle stage and root cause.

\noindent\begin{minipage}{\columnwidth}
  \setlength{\belowcaptionskip}{5pt}
  \centering
  \setlength{\belowcaptionskip}{3pt}
  \refstepcounter{table}
  {\small\textbf{Table~\thetable:} Container Path Traversal Vulnerabilities\par}
  \label{tab:path-traversal-vulnerabilities}
  \scriptsize
  \setlength{\tabcolsep}{3pt}
  \renewcommand{\arraystretch}{1.02}

  \begin{tabular}{llll}
    \toprule
    CVE ID & Component & Lifecycle & Root Cause \\
    \midrule
    CVE-2017-1002101 & Kubernetes & Pre-Start & TOCTOU \\
    \midrule
    CVE-2018-15664 & Docker & Runtime & TOCTOU \\
    \midrule
    CVE-2019-10152 & Podman & Runtime & MisResolution \\
    CVE-2019-16884 & runc & Pre-Start & MisResolution \\
    CVE-2019-18466 & Podman & Pre-Start & MisResolution \\
    CVE-2019-19921 & runc & Pre-Start & TOCTOU \\
    \midrule
    CVE-2020-2024 & Kata Containers & Runtime & MisResolution \\
    CVE-2020-2026 & Kata Containers & Pre-Start & MisResolution \\
    CVE-2020-8567 & Kubernetes & Pre-Start & MisResolution \\
    CVE-2020-10696 & Buildah & Pre-Start & MisResolution \\
    \midrule
    CVE-2021-25741 & Kubernetes & Runtime & TOCTOU \\
    CVE-2021-30465 & runc & Runtime & TOCTOU \\
    \midrule
    CVE-2022-23648 & Containerd & Pre-Start & MisResolution \\
    \midrule
    CVE-2023-0778 & Podman & Runtime & TOCTOU \\
    CVE-2023-27561 & runc & Pre-Start & TOCTOU \\
    CVE-2023-28642 & runc & Pre-Start & MisResolution \\
    CVE-2023-37274 & Auto-GPT & Runtime & MisResolution \\
    \midrule
    
    CVE-2024-0132 & NVIDIA & Pre-Start & TOCTOU \\
    CVE-2024-0133 & NVIDIA & Pre-Start & TOCTOU \\
    CVE-2024-0134 & NVIDIA & Pre-Start & MisResolution \\
    CVE-2024-5154 & CRI-O & Runtime & MisResolution \\
    CVE-2024-11218 & Buildah & Pre-Start & TOCTOU \\
    CVE-2024-23651 & Buildkit & Pre-Start & TOCTOU \\
    CVE-2024-23652 & Buildkit & Pre-Start & TOCTOU \\
    \midrule
    
    CVE-2025-9566 & Podman & Pre-Start & MisResolution \\
    CVE-2025-23267 & NVIDIA & Pre-Start & MisResolution \\
    CVE-2025-23359 & NVIDIA & Pre-Start & TOCTOU \\
    CVE-2025-24965 & crun & Pre-Start & MisResolution \\
    CVE-2025-31133 & runc & Pre-Start & TOCTOU \\
    CVE-2025-47290 & Containerd & Pre-Start & TOCTOU \\
    CVE-2025-52565 & runc & Pre-Start & TOCTOU \\
    CVE-2025-52881 & runc & Pre-Start & TOCTOU \\
    \midrule
    
    CVE-2026-7875 & NanoClaw & Runtime & MisResolution \\
    CVE-2026-28406 & Kaniko & Pre-Start & MisResolution \\
    CVE-2026-31990 & OpenClaw & Runtime & MisResolution \\
    CVE-2026-41568 & Moby & Runtime & TOCTOU \\
    CVE-2026-42306 & Moby & Runtime & TOCTOU \\
    CVE-2026-44112 & OpenClaw & Runtime & TOCTOU \\
    CVE-2026-44113 & OpenClaw & Runtime & TOCTOU \\
    CVE-2026-46703 & Boxlite & Pre-Start & MisResolution \\
    CVE-2026-54319 & Daytona & Pre-Start & MisResolution \\
    CVE-2026-56692 & NanoClaw & Runtime & MisResolution \\
    CVE-2026-55686 & Podman & Pre-Start & MisResolution \\
    CVE-2026-41579 & runc & Pre-Start & MisResolution \\
    CVE-2026-53489 & Containerd & Pre-Start & MisResolution \\
    CVE-2026-24260 & NVIDIA & Pre-Start & TOCTOU \\
    CVE-2026-17106 & Moby & Runtime & MisResolution \\
    \bottomrule
  \end{tabular}
\end{minipage}

\section{Formal Templates and Prompt Contracts}
\label{app:formal-templates-prompts}

Bulkhead separates candidate generation from the definition of a valid repair.
For a formal trial, the runner pins the vulnerable Git revision and hash-binds
the generation bundle, an accepted independently reviewed mapping, and the
selected template bytes. The template fixes the finite state domain, attacker behavior, initialization,
security property, and compatibility assertion. The LLM cannot redefine these
elements.

The following listings are contract excerpts rather than verbatim prompt dumps.
The anonymous artifact contains the complete versioned prompts and templates.
The runner enforces the interface independently of the prompt wording.

\begin{table}[t]
  \centering
  \setlength{\belowcaptionskip}{2.5pt}
  \scriptsize
  \setlength{\tabcolsep}{3pt}
  \renewcommand{\arraystretch}{1.05}
  \caption{Data and control boundaries in remediation.}
  \label{tab:prompt-metavariables}
  \begin{tabularx}{\columnwidth}{@{}p{0.19\columnwidth}>{\raggedright\arraybackslash}X@{}}
    \toprule
    \textbf{Owner} & \textbf{Bound interface} \\
    \midrule
    Runner &
    Pinned vulnerable revision, template-owned state declarations and finite
    domains, fixed interleavings, attacker process, initialization, properties,
    and compatibility assertion. \\
    Patch LLM &
    Receives the attack summary, authenticated source windows, call chain,
    build target, assumptions, and atomic predicates; returns one unified diff. \\
    Translation LLM &
    Returns three declared Promela statement snippets and one source binding
    for each snippet. \\
    Withheld &
    Full@3 excludes the reference repair, repaired revision, evaluator commands,
    hidden harness details, and terminal evaluation outcomes. \\
    \bottomrule
  \end{tabularx}
\end{table}

\subsection{Pattern-Template Contract}
\label{app:pattern-template-family}

The matched lifecycle and root cause select one human-authored template.
\texttt{pwc} is the artifact identifier retained for the MisResolution class.
The four cases are listed in Table~\ref{tab:pattern-template-family}.

\begin{table}[t]
  \centering
  \scriptsize
  \setlength{\tabcolsep}{3pt}
  \renewcommand{\arraystretch}{1.05}
  \caption{Hash-bound template family.}
  \label{tab:pattern-template-family}
  \begin{tabular}{@{}lcc@{}}
    \toprule
    & \textbf{MisResolution} & \textbf{TOCTOU} \\
    \midrule
    Pre-Start & \texttt{pre\_startup\_pwc} & \texttt{pre\_startup\_toctou} \\
    Runtime & \texttt{runtime\_pwc} & \texttt{runtime\_toctou} \\
    \bottomrule
  \end{tabular}
\end{table}

MisResolution templates model a path planted before a privileged operation.
TOCTOU templates preserve the relevant check-to-use interleavings. Each
template exercises an escaping symlink, an in-boundary symlink, and a normal
path. Listing~\ref{lst:prestartup-misresolution-template} shows the structural
interface. The runner supplies \texttt{init\_state} from the hash-bound
template. Translation fills only the three marked transition slots.

\begin{lstlisting}[
  style=bulkheadappendix,
  caption={Structural excerpt of the Pre-Start MisResolution template.},
  label={lst:prestartup-misresolution-template}
]
#define NORMAL          0
#define SYMLINK_INSIDE  1
#define SYMLINK_OUTSIDE 2

byte current_file_type = NORMAL;
byte attacker_mode = 0;
bit host_escaped = 0;
bit op_completed = 0;

ltl prestartup_pwc_safe {
  [] (host_escaped == 0)
}

proctype HostUtility() {
  setup_trigger?1;
  // __VALIDATION_SEQUENCE__
  // __OPERATION_SEQUENCE__
  // __INVARIANT_CHECK__
  assert(attacker_mode != 1 ||
         op_completed == 1);
}

init {
  // __INIT_STATE__  /* runner-owned */
  atomic {
    if
    :: attacker_mode = 0;
       current_file_type = SYMLINK_OUTSIDE
    :: attacker_mode = 1;
       current_file_type = SYMLINK_INSIDE
    :: attacker_mode = 2;
       current_file_type = NORMAL
    fi;
    run HostUtility();
    setup_trigger!1
  }
}
\end{lstlisting}

The LTL property rejects boundary escape. The assertion preserves legitimate
behavior: attacker mode~1 represents an in-boundary symlink for which the
operation must complete. A patch that rejects every symlink may satisfy the
security property but fails this compatibility check.

\subsection{Generation and Translation Contracts}
\label{app:prompt-contracts}

Patch synthesis and candidate translation use separate interfaces. The first
produces repository code. The second maps the effects of that code into the
already selected template.

\begin{lstlisting}[
  style=bulkheadappendix,
  caption={Patch synthesis contract excerpt.},
  label={lst:patch-generation-prompt}
]
BULKHEAD PATCH GENERATION PROMPT
PROMPT VERSION: bulkhead-patch-v1

VISIBLE INPUT
  case, repository, vulnerable revision
  build target and attack summary
  authenticated source windows
  entry-to-effect call chain
  reviewed attacker model, assumptions,
  and atomic security predicates

REQUIRE
  complete minimal diff that applies and builds
  preserve behavior allowed by fixed context
  do not alter context or invent source facts

OUTPUT
  one complete raw unified diff; no prose
\end{lstlisting}

\begin{lstlisting}[
  style=bulkheadappendix,
  caption={Candidate-to-model contract excerpt.},
  label={lst:model-translation-prompt}
]
BULKHEAD MODEL INSTANTIATION
PROMPT VERSION:
  bulkhead-model-translation-v1

INPUT
  candidate diff and frozen formal context

OUTPUT: raw JSON with exactly
  injections:
    three declared statement slots
  source_bindings:
    one record per statement slot

EACH BINDING
  identifies one distinct complete added line
  names its frozen predicate IDs

REJECT
  init_state or extra keys
  empty, vacuous, or structural Promela
  unconditional safety assignments
\end{lstlisting}

The prompt is not the enforcement mechanism. The runner checks the exact key
set, rejects empty or structural snippets, and requires each snippet to
reference a bound frozen predicate. Every binding must name a distinct exact added-diff-line excerpt; that excerpt
must occur exactly once in the candidate diff. Invalid
translations are rejected before SPIN runs. These checks prevent the translator
from redefining the oracle, but do not by themselves prove source-to-model
equivalence.

The frozen \texttt{Full@3} protocol permits one initial candidate and at most
two repair candidates. Each repair request contains the previous complete diff
and one bounded apply, build, or model-checking diagnostic.

\begin{lstlisting}[
  style=bulkheadappendix,
  caption={Typed repair-feedback contract excerpt.},
  label={lst:repair-feedback-prompt}
]
PREVIOUS CANDIDATE
  <complete unified diff>

FEEDBACK CATEGORY
  PATCH_APPLY_FAILED
  BUILD_FAILED
  MODEL_COUNTEREXAMPLE

DIAGNOSTIC
  <bounded public excerpt>

OUTPUT
  <new complete raw unified diff>
\end{lstlisting}

The same protocol reserves runtime replay, benign controls, and sealed
confirmation for terminal scoring. Their outcomes do not enter repair prompts.

\subsection{Interpretation of Verification Outcomes}
\label{app:verification-boundary}

The gates support distinct claims. Successful patch application and a passing
build show only that a candidate can be integrated. A completed SPIN exploration with no
counterexample supports the security and compatibility properties only within
the selected finite abstraction and its fixed attacker behavior. Invalid model
instantiations, incomplete searches, timeouts, and verifier errors do not count
as safe. Runtime exploit replay, benign controls, and sealed confirmation
evaluate the concrete implementation separately. A repair is fully validated
only when every required gate succeeds; the model-checking result does not
establish the absence of vulnerabilities outside the modeled state space.